\documentclass[superscriptaddress,longbibliography,aps,prb,reprint,floatfix,preprintnumbers]{revtex4-2}
\usepackage{graphicx}
\usepackage[dvipsnames]{xcolor}
\usepackage{subfigure}
\usepackage{amssymb,amsmath,mathptmx}
\usepackage[colorlinks=true,citecolor=blue,linkcolor=blue,urlcolor=blue,bookmarks=false]{hyperref}
\usepackage{tikz}
\usepackage[T1]{fontenc}
\usepackage[utf8]{inputenc}
\usepackage{CJKutf8}

\graphicspath{fig/}

\DeclareMathOperator{\sgn}{sgn}

\def\unit#1{\mathord{\thinspace\rm #1}}

\begin{document}
\begin{CJK*}{UTF8}{bsmi}

\title{Scalable tight-binding model for strained graphene}

\author{Ming-Hao Liu (劉明豪)}

\email{minghao.liu@phys.ncku.edu.tw}

\affiliation{Department of Physics and Center for Quantum Frontiers of Research and Technology (QFort), National Cheng Kung University, Tainan 70101, Taiwan}

\author{Christophe De Beule}
\email{christophe.debeule@uantwerpen.be}
\affiliation{Department of Physics and Astronomy, University of Pennsylvania, Philadelphia, Pennsylvania 19104, USA}
\affiliation{Department of Physics, University of Antwerp, Groenenborgerlaan 171, 2020 Antwerp, Belgium}

\author{Alina Mre\'nca-Kolasi\'nska}
%\author{\protect\\ Alina Mre\'nca-Kolasi\'nska}
%\email{alina.mrenca@fis.agh.edu.pl}
\affiliation{AGH University of Krakow, Faculty of Physics and Applied Computer Science, al. Mickiewicza 30, 30-059 Krak\'ow, Poland}

\author{Hsin-You~Wu~(吳欣祐)}

\affiliation{Department of Physics and Center for Quantum Frontiers of Research and Technology (QFort), National Cheng Kung University, Tainan 70101, Taiwan}

\author{Aitor Garcia-Ruiz (艾飛宇)}

\affiliation{Department of Physics and Center for Quantum Frontiers of Research and Technology (QFort), National Cheng Kung University, Tainan 70101, Taiwan}

\author{Denis Kochan}

\email{denis.kochan@phys.ncku.edu.tw}

\affiliation{Department of Physics and Center for Quantum Frontiers of Research and Technology (QFort), National Cheng Kung University, Tainan 70101, Taiwan}

\author{Klaus Richter}

%\affiliation{Institut f\"ur Theoretische Physik, Universit\"at Regensburg, D-93040 Regensburg, Germany}

\affiliation{Institute for Theoretical Physics and Halle-Berlin-Regensburg Cluster of Excellence CCE, University of Regensburg, 93040 Regensburg, Germany}

\date{\today}

\preprint{\href{https://journals.aps.org/prb/abstract/10.1103/ntf2-5dyw}{Phys.\ Rev.\ B \textbf{113}, 195429} -- Published 18 May, 2026}

\begin{abstract}
We generalize the scalable tight-binding model for graphene, which allows for efficient quantum transport simulations in the Dirac regime, to account for elastic strain. We show that the original scalable model with scaling factor $s$ is readily applicable to strained graphene, provided that the displacement fields corresponding to the deformed graphene lattice are properly scaled. In particular, we show that the long-wavelength theory remains invariant when the strain tensor is scaled by $s$. This is achieved in practice by scaling the in-plane displacement fields by $s$ while the out-of-plane displacements have to be scaled by $\sqrt{s}$. We confirm these scaling laws by extensive numerical simulations, starting with the pseudomagnetic field and the local density of states for different scaled lattices. The latter allows us to study pseudo-Landau levels as well as hybrid Landau levels in the presence of an external magnetic field. Finally, we consider quantum transport simulations motivated by a recent experiment, where a uniaxial strain barrier is engineered in monolayer graphene by vertically misaligned gates. Our work generalizes the scalable tight-binding model to allow for efficient modeling of quantum transport in large-scale strained graphene devices.
\end{abstract}

\maketitle

\end{CJK*}

\section{Introduction}

More than two decades ago, the era of two-dimensional materials was ushered in with the experimental discovery of graphene \cite{Novoselov2004}. Since then, graphene has served as an ideal testbed for electron optics, owing to its unusual linear-in-momentum energy dispersion for low-energy charge carriers that mimics relativistic Dirac fermions \cite{CastroNeto2009}. The development of ultraclean graphene devices through advanced fabrication techniques has been decisive in revealing novel transport phenomena 
\cite{Chakraborti2024}. On the theory side, large-scale quantum transport simulations are important to establish direct correspondence between experiment and theory. In part, the latter has been made possible by the scalable tight-binding model (TBM) introduced about a decade ago \cite{Liu2015}.

\begin{figure}[!b]
    \centering
    \includegraphics[width=\columnwidth]{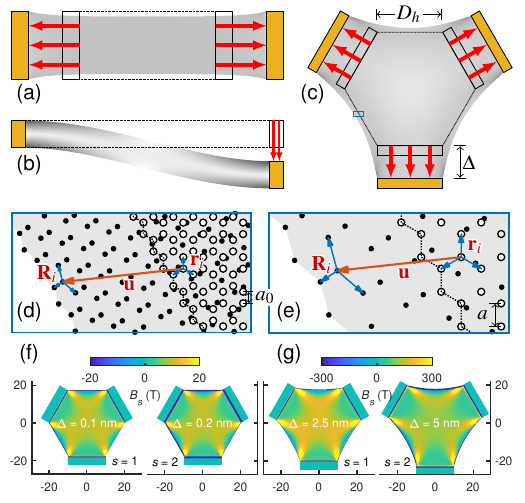}
    \caption{Examples of strained graphene devices: (a) uniaxial strain applied to a graphene ribbon, (b) shear strain applied to a graphene ribbon with one end fixed, and (c) triaxial strain applied to a hexagonal graphene flake. The region marked by a small blue rectangle in (c) is magnified in (d) and (e), considering an unscaled ($s=1$) and scaled ($s=2$) graphene lattice, respectively, \textit{without} scaling the in-plane displacement field $\mathbf u$, where empty (solid) dots represent lattice points in the graphene sheet before (after) the strain is applied. (f) Calculated PMF profiles of a $D_h=20\unit{nm}$ flake considering $s=1$ with $\Delta=0.1\unit{nm}$ (left) and $s=2$ with $\Delta=0.2\unit{nm}$ (right); $D_h$ and $\Delta$ are defined in (c). (g) Same as (f) but with a larger $\Delta$ as indicated.}
    \label{fig schematics}
\end{figure}

In this approach, the carbon-carbon distance $a_0$ is scaled up by a scaling factor $s > 1$, while the nearest-neighbor hopping strength $t_0$ is reduced by the same factor to $t_0/s$, ensuring that the underlying long-wavelength Dirac theory remains invariant as long as any external fields vary slowly on the scaled lattice. In this way, no information is lost when the fields are sampled on the scaled lattice. The main advantage of the scaling method is that the Hamiltonian matrix size is reduced by a factor $1/s^2$. For example, for a single-orbital spinless tight-binding description of graphene, the number of entries is reduced from 38 million by 38 million per micron squared to $38/s^2$ million by $38/s^2$ million. Depending on the energy range (or equivalently, the carrier density range) that one is interested in, the scaling factor can be tuned accordingly, limited by $\lambda_F \gg sa_0$. As a concrete example, a $1\unit{\mu m}\times 1\unit{\mu m}$ graphene flake scaled by $s=10$ allows a reliable tight-binding description up to $\sim\pm 0.28\unit{eV}$ corresponding to nearly $\pm 6\times 10^{12}\unit{cm^{-2}}$ of carrier densities using only $0.38$ million of lattice sites, which is computationally fine nowadays. Without scaling, on the other hand, a lattice of $38$ million lattice sites is still intractable today. Reducing the number of lattice sites leads to downsizing the dimension of the Hamiltonian matrix, which is the main computational obstacle for quantum transport simulations working in real space. For more specific applications of the scalable TBM to electron-optics experiments, we refer to Ref.\ \onlinecite{Chakraborti2024} and the references therein.

Beyond the optics-like transport properties of pristine graphene, mechanically deformed graphene, which we refer to as strained graphene, offers a versatile platform for exploring pseudomagnetotransport, i.e., transport under strain-engineered pseudomagnetic fields (PMFs). Such deformations can be achieved either by depositing graphene on suitable substrates \cite{Zhang2018a,Mao2020,Kang2021,Milovanovic2019,Milovanovic2020} or by applying external stress \cite{Guinea2009,Zhu2015,Androulidakis2015}. Representative examples include uniaxially strained graphene \cite{Pereira2009}, bent graphene ribbons \cite{Guinea2010,Kapfer2023}, and triaxially strained graphene \cite{Guinea2009,NeekAmal2013}, schematically illustrated in Fig.\ \ref{fig schematics}(a), (b), and (c), respectively. The PMF is given by $\mathbf{B}_s=\nabla\times\mathbf{A}_s$ where $\mathbf{A}_s$ is the intravalley pseudogauge field due to shear strain that varies slowly on the graphene lattice scale \cite{Kane1997,Suzuura2002,Katsnelson2007,Vozmediano2010,Amorim2016}. In a nutshell, shear strain breaks the intrinsic $120^\circ$ rotation symmetry about a carbon bond, shifting the Dirac cone in momentum space away from the corners of the hexagonal Brillouin zone. Hence, shear strain couples to the momentum in the low-energy Dirac Hamiltonian in the same way as minimal coupling, $\mathbf{p}\rightarrow \mathbf{p} \pm e\mathbf{A}_s$, where the plus and minus signs apply to valleys $\mathbf{K}$ and $\mathbf{K}'$, preserving the time reversal symmetry of the deformed lattice. Despite long-standing theoretical predictions, experimental investigations of pseudomagnetotransport in strained graphene remain limited \cite{Zhang2018a,Zhang2019,Zhang2022}. This limited progress reflects not only the formidable experimental challenges but also the absence of efficient quantum transport methods for large-scale strained graphene. 

In this paper, we show that the scalable TBM can be readily applied to strained graphene, on the condition that the displacement fields corresponding to the atomic deformations in the strained graphene lattice are properly scaled:
\begin{align} \label{eq:scaling}
    a_0 \rightarrow s a_0\ , && t_0 \rightarrow t_0/s\ , && \mathbf u \rightarrow s \mathbf u\ , && h \rightarrow \sqrt{s} h\ ,
\end{align}
where $\mathbf{u}$ and $h$ are in-plane and out-of-plane displacement fields, respectively. The latter are introduced in Section \ref{sec scalable TBM for strained graphene}, where we also provide an intuitive picture using Fig.\ \ref{fig schematics}(d) and (e) to explain why the displacements need to be scaled. As a quick numerical visualization, Fig.\ \ref{fig schematics}(f) shows how well the scaling works when the strain is small. However, scaling may fail when the displacement field is too large, as exemplified in  Fig.\ \ref{fig schematics}(g). More details are given in Section \ref{sec lattice pmf}.

The generalized scalable TBM described by Eq.\ \eqref{eq:scaling} enables quantum transport simulations for strained graphene at experimentally relevant scales, and is the main message of this paper, which is organized as follows. In Section \ref{sec theory}, we introduce the TBM for graphene, starting with a scaled lattice without strain (Section \ref{sec TBM for unstrained graphene}), and strained graphene without scaling (Section \ref{sec TBM for strained graphene}), both being well established in the literature. In Section \ref{sec scalable TBM for strained graphene}, we then derive the scaling law for the acoustic displacement field, i.e., the latter part of Eq.\ \eqref{eq:scaling}. In Section \ref{sec numerics}, we present numerical simulations to verify our generalized scalable TBM, including lattice PMF (Section \ref{sec lattice pmf}), local density of states (mainly Section \ref{sec ldos} with additional data in Appendix \ref{appendix additional data}), and transport simulations (Section \ref{sec transport}). We present our conclusions in Section \ref{sec summary}.

\section{Theory}\label{sec theory}

The main purpose of this section is to derive the scaling law of the displacement field summarized in Eq.\ \eqref{eq:scaling}. To this end, we start with a short review of the standard TBM for scaled graphene without strain and strained graphene without scaling. At the end of this section, we also provide an intuitive picture explaining why the displacement field must be scaled when implementing the scalable TBM.

\subsection{TBM for unstrained graphene}\label{sec TBM for unstrained graphene}

We start with the tight-binding Hamiltonian describing nearest-neighbor hopping between carbon $p_z$ orbitals in graphene,
\begin{equation}
\label{eq:Htb}
H = \sum\limits_i {U({\mathbf{r}}_i)} c_i^\dagger c_i -\sum\limits_{\left\langle {i,j} \right\rangle } t_{ij} c_i^\dagger  c_j\ ,
\end{equation}
where $c_i$ ($c_i^\dag$) annihilates (creates) an electron on site $i$ at position $\mathbf{r}_i = (x_i,y_i)$. The first sum on the left-hand side that runs over all sites describes the onsite potential energy $U$, which accounts for gate electrodes, potential disorder, chemical doping, and so on.

In the \textit{absence} of strain, the double sum in Eq.\ \eqref{eq:Htb} runs over all nearest neighbors, i.e., pairs of sites $i,j$ such that $|\mathbf{r}_i-\mathbf{r}_j|=a_0$ and $t_{ij}=t_0$ for an unscaled graphene lattice. We adopt tight-binding parameters $a_0 = (4\sqrt{3})^{-1}\unit{nm}$ and $t_0=3\unit{eV}$ throughout this paper. When the graphene lattice is scaled by $s$, the tight-binding parameters are scaled to $a_0 \rightarrow sa_0$ and $t_0 \rightarrow t_0/s$, which are the first part of Eq.\ \eqref{eq:scaling} and explained in Ref.\ \onlinecite{Liu2015}.

\subsection{TBM for strained graphene without scaling}\label{sec TBM for strained graphene}

For monolayer graphene with applied elastic strain, the tight-binding Hamiltonian \eqref{eq:Htb} without scaling remains in the same form, but the position of the $i$th site may be displaced from $\mathbf{r}_i$ (before strain) to $\mathbf{R}_i$ (after strain) by an acoustic displacement field $\mathbf{u}(x_i,y_i) + h(x_i,y_i) \hat{\mathbf{e}}_z$ that acts the same on both sublattices, where $\mathbf u(x,y)$ is the in-plane component and $h(x,y)$ is the out-of-plane component, $\hat{\mathbf{e}}_z$ being the unit vector along the $z$ axis (and similarly $\hat{\mathbf{e}}_x$ and $\hat{\mathbf{e}}_y$). That is,
\begin{equation}
    \mathbf{R}_i = \mathbf{r}_i + \mathbf{u}(\mathbf r_i) + h(\mathbf r_i) \hat{\mathbf{e}}_z\ ,
\label{eq def r=r0+u+h}
\end{equation}
where we use a Lagrangian description. This is illustrated in Fig.\ \ref{fig schematics}(d), taking Fig.\ \ref{fig schematics}(c) under an in-plane strain as an example. For corrugations that vary slowly with respect to the graphene lattice spacing, such that local curvature effects \cite{Kim2008} can be neglected ($a \nabla^2 h \ll 1$), the hopping amplitude can be modeled in the central-force approximation \cite{Pereira2009} as
\begin{equation}
    t_{ij} = t(d_{ij}) = t_0 \exp\left[-\beta\left(\frac{d_{ij}}{a_0}-1\right)\right]\ ,
\label{eq tij}
\end{equation}
where $d_{ij} = |\mathbf{R}_i-\mathbf{R}_j|$ is the distance between sites labeled by $i$ and $j$, and the electron Gr\"uneisen parameter, $\beta \equiv -(a_0/t_0) \left. \partial t/\partial d \right|_{d=a_0}$, is taken to be $\beta = 3.37$ \cite{Pereira2009}.

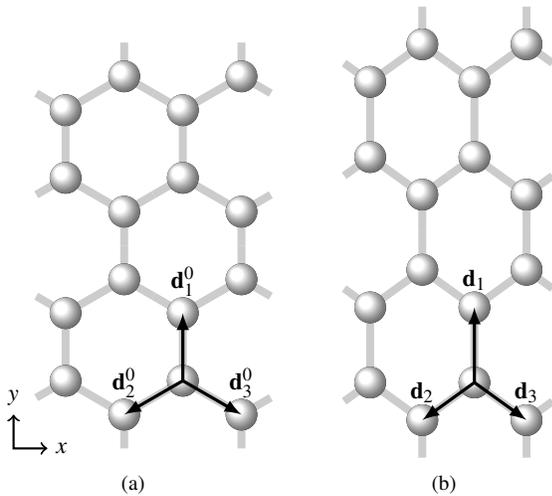
\begin{figure}[b]

\subfigure[\label{fig graphene nn vectors}]{
\begin{tikzpicture}
\def\bondColor{black!20}
\def\ballColor{white}
\def\dotSize{6.6667}

\begin{scope}[line width=3pt,scale=0.9]

\draw[<->,thick] (-2.5,-1) node [above] {$y$} |- ++ (0.5,-0.5) node [right] {$x$};

\foreach \ny in {0,1}{
\foreach \nx in {-1,0}{
\begin{scope}[xshift=sqrt(3)*1cm*\nx,yshift=3*1cm*\ny]
%%%% upper zigzag
\draw [\bondColor] ({-sqrt(3)/4},3/4) -- (0,1/2) -- ({sqrt(3)/2},1) -- ({3/4*sqrt(3)},3/4);
%%%% lower zigzag
\draw [\bondColor] ({-sqrt(3)/4},-3/4) -- (0,-1/2) -- ({sqrt(3)/2},-1) -- ({3/4*sqrt(3)},-3/4);
%%%% vertical lines
\draw [\bondColor] (0,1/2) -- (0,-1/2);
\draw [\bondColor] ({sqrt(3)/2},1) -- +(0,1/2);
\draw [\bondColor] ({sqrt(3)/2},-1) -- +(0,-1/2);
%%%% sublattice sites: A
\fill [shading=ball,ball color=\ballColor] ({sqrt(3)/2},1) circle (\dotSize*1pt);
\fill [shading=ball,ball color=\ballColor] (0,-1/2) circle (\dotSize*1pt);
%%%% sublattice sites: B
\fill [shading=ball,ball color=\ballColor] ({sqrt(3)/2},-1) circle (\dotSize*1pt);
\fill [shading=ball,ball color=\ballColor] (0,1/2) circle (\dotSize*1pt);
\end{scope}
}}

%%%% bond vectors
\draw [very thick,-latex] (0,-0.5) -- ++(90:1) node [above=3pt] {$\mathbf{d}_1^0$};
\draw [very thick,-latex] (0,-0.5) -- ++(210:1) node [above=3pt] {$\mathbf{d}_2^0$};
\draw [very thick,-latex] (0,-0.5) -- ++(-30:1) node [above=3pt] {$\mathbf{d}_3^0$};
\end{scope}
\end{tikzpicture}
}
\qquad
\subfigure[\label{fig graphene nn vectors strained}]{
\begin{tikzpicture}
\def\bondColor{black!20}
\def\ballColor{white}
\def\dotSize{6}

\begin{scope}[line width=3pt]
\foreach \ny in {0,1}{
\foreach \nx in {-1,0}{
\begin{scope}[xshift=sqrt(3)*0.8cm*\nx,yshift=3*1cm*\ny]
%%%% upper zigzag
\draw [\bondColor] ({-sqrt(3)/4*0.8},3/4) -- (0,1/2) -- ({sqrt(3)/2*0.8},1) -- ({3/4*sqrt(3)*0.8},3/4);
%%%% lower zigzag
\draw [\bondColor] ({-sqrt(3)/4*0.8},-3/4) -- (0,-1/2) -- ({sqrt(3)/2*0.8},-1) -- ({3/4*sqrt(3)*0.8},-3/4);
%%%% vertical lines
\draw [\bondColor] (0,1/2) -- (0,-1/2);
\draw [\bondColor] ({sqrt(3)/2*0.8},1) -- +(0,1/2);
\draw [\bondColor] ({sqrt(3)/2*0.8},-1) -- +(0,-1/2);
%%%% sublattice sites: A
\fill [shading=ball,ball color=\ballColor] ({sqrt(3)/2*0.8},1) circle (\dotSize*1pt);
\fill [shading=ball,ball color=\ballColor] (0,-1/2) circle (\dotSize*1pt);
%%%% sublattice sites: B
\fill [shading=ball,ball color=\ballColor] ({sqrt(3)/2*0.8},-1) circle (\dotSize*1pt);
\fill [shading=ball,ball color=\ballColor] (0,1/2) circle (\dotSize*1pt);
\end{scope}
}}

%%%% bond vectors
\draw [very thick,-latex] (0,-0.5) -- ++(90:1) node [above=3pt] {$\mathbf{d}_1$};
\draw [very thick,-latex] (0,-0.5) -- ++({-sqrt(3)/2*0.8},-1/2) node [above=3pt] {$\mathbf{d}_2$};
\draw [very thick,-latex] (0,-0.5) -- ++({sqrt(3)/2*0.8},-1/2) node [above=3pt] {$\mathbf{d}_3$};
\end{scope}
\end{tikzpicture}
}

\caption{Illustration of (a) an unstrained graphene lattice with nearest-neighbor bond vectors $\{ \mathbf{d}_1^0, \mathbf{d}_2^0, \mathbf{d}_3^0 \}$ and (b) a strained graphene lattice with nearest-neighbor bond vectors $\{ \mathbf{d}_1, \mathbf{d}_2, \mathbf{d}_3 \}$.}

\end{figure}

\subsection{Scalable TBM for strained graphene}\label{sec scalable TBM for strained graphene}

To account for strain in the scalable TBM, we first expand the hopping function given in Eq.\ \eqref{eq tij} as $t_{ij} = t_0 + \delta t_{ij}$, where $\delta t_{ij} \approx -\beta t_0 (d_{ij}/a_0 -1)$. If we retain only nearest-neighbor hopping, the pseudo vector potential from Eq.\ \eqref{eq:As} in Appendix \ref{appendix pseudogauge fields} (where a brief overview of the pseudogauge field is given) becomes
\begin{equation}
%t_0 \delta t_{ij} 
    A_{s,x} - i A_{s,y} = -\frac{1}{ev_F} \sum\limits_{n=1}^3 \delta t_n e^{i \mathbf{K} \cdot \mathbf d_n^0} \ ,
    \label{eq:As_perturb}
\end{equation}
where $v_F = 3t_0a_0 / (2\hbar)$ is the Fermi velocity of graphene and $\delta t_n(\mathbf r)$ is the change in the nearest-neighbor hopping amplitude evaluated at the undeformed sublattice A, and the pristine nearest-neighbor bond vectors $\mathbf d_n^0$ are shown in Fig.\ \ref{fig graphene nn vectors}. Placing the $x$ axis along the zigzag direction, we can take $\mathbf{K} = 4\pi / (3\sqrt{3}a_0) \hat{\mathbf{e}}_x$ and Eq.\ \eqref{eq:As_perturb} gives \cite{Katsnelson2007,Vozmediano2010}
\begin{equation}
\begin{aligned}
    A_{s,x} & = -\frac{1}{2ev_F} \left( 2 \delta t_1 - \delta t_2 + \delta t_3 \right ) \\
    A_{s,y} & =  -\frac{\sqrt{3}}{2ev_F} \left( \delta t_2 - \delta t_3 \right ).
\label{eq As final}
\end{aligned}
\end{equation}
Note that Eq.\ \eqref{eq As final} can be used to compute lattice-site-resolved PMF, as described later in Section \ref{sec lattice pmf}.

In the lowest order of displacements,
%the change in the hopping becomes
\begin{equation}
    \delta t_{n} = -\frac{\beta t_0}{a_0^2} \left( \mathbf{d}_n^0 \cdot (\mathbf{d}_{n}-\mathbf{d}_{n}^0) + \frac{1}{2} \left[ (\mathbf{d}_{n}-\mathbf{d}_{n}^0)_z \right]^2 \right), \label{eq:dt}
\end{equation}
where the deformed bond vectors $\mathbf{d}_n = (d_{nx}, d_{ny}, d_{nz})$ are exemplified in Fig.\ \ref{fig graphene nn vectors strained} for in-plane uniaxial strain along the armchair direction. The components of $\mathbf{A}_\mathrm{s}$ become
\begin{equation}
% h is reserved for the displacement field itself, not the z-component of the bond vector
\begin{aligned}
    % A_{s,x} =&\   \frac{\beta t_0}{2ev_F} \left[ 2 \left(\frac{d_{1y}}{a_0} - 1 \right) - \left(\frac{-\sqrt{3} d_{2x} - d_{2y}}{2a_0} - 1\right) \right. \\ 
    % & - \left.  \left(\frac{\sqrt{3} d_{3x} - d_{3y}}{2a_0} - 1\right) + \frac{2h_1^2-h_2^2-h_3^2}{a_0^2} \right ] \\
    % A_{s,y} = &\  \frac{\beta\sqrt{3}t_0}{2ev_F} \left[\left(\frac{-\sqrt{3}d_{2x}-d_{2y}}{2a_0}-1\right) \right.\\ 
    % & -\left. \left(\frac{\sqrt{3} d_{3x}-d_{3y}}{2a_0} -1\right) + \frac{h_2^2-h_3^2}{a_0^2} \right]. \label{eq:As_xy}
    A_{s,x} =&\   \frac{\beta t_0}{2ev_F} \left[ 2 \left(\frac{d_{1y}}{a_0} - 1 \right) - \left(\frac{-\sqrt{3} d_{2x} - d_{2y}}{2a_0} - 1\right) \right. \\ 
    & - \left.  \left(\frac{\sqrt{3} d_{3x} - d_{3y}}{2a_0} - 1\right) + \frac{2d_{1z}^2-d_{2z}^2-d_{3z}^2}{a_0^2} \right ] \\
    A_{s,y} = &\  \frac{\beta\sqrt{3}t_0}{2ev_F} \left[\left(\frac{-\sqrt{3}d_{2x}-d_{2y}}{2a_0}-1\right) \right.\\ 
    & -\left. \left(\frac{\sqrt{3} d_{3x}-d_{3y}}{2a_0} -1\right) + \frac{d_{2z}^2-d_{3z}^2}{a_0^2} \right], \label{eq:As_xy}
\end{aligned}
\end{equation}
where we set $d_{nz}^0 = 0$. We now consider how the pseudogauge field transforms under the original scaling transformation \cite{Liu2015}. When the lattice is scaled by a factor $s$ and the hopping $t_0$ with a factor $1/s$, we see that Eq.\ \eqref{eq:dt} remains invariant if
\begin{alignat}{3}
    & ( \mathbf{d}_{n}-\mathbf{d}_{n}^0 )_{x,y} && \rightarrow s^2  && ( \mathbf{d}_{n}-\mathbf{d}_{n}^0 )_{x,y}, \\
    & ( \mathbf{d}_{n}-\mathbf{d}_{n}^0 )_z && \rightarrow s^{3/2} && ( \mathbf{d}_{n}-\mathbf{d}_{n}^0 )_z.
\end{alignat}
Moreover, in the absence of optical displacements
\begin{align}
    \mathbf d_n(\mathbf r) - \mathbf d_n^0 & = \mathbf u(\mathbf r + \mathbf d_n^0) - \mathbf u(\mathbf r) + \left[ h(\mathbf r + \mathbf d_n^0) - h( \mathbf r) \right] \hat{\mathbf e}_z \nonumber \\
    & \approx ( \mathbf d_n^0 \cdot \nabla ) \left[ \mathbf u(\mathbf r) + h(\mathbf r) \hat{\mathbf e}_z \right]. \label{eq:linear}
\end{align}
Thus, as long as the linear expansion in Eq.\ \eqref{eq:linear} holds, the in-plane displacement field has to be scaled by an additional factor $s$. Similarly, out-of-plane displacements have to be scaled by a factor $s^{1/2}$ because they enter at quadratic order in Eq.\ \eqref{eq:dt}. These conclude with $\mathbf u \rightarrow s \mathbf u$ and $h \rightarrow \sqrt{s} h$, which are the latter part of Eq.\ \eqref{eq:scaling}.

While we only explicitly consider acoustic degrees of freedom $\mathbf u = (\mathbf u_\text{A} + \mathbf u_\text{B})/2$ and similar for out-of-plane displacements, contributions from optical displacements $\mathbf v = \mathbf u_\text{A} - \mathbf u_\text{B}$ are accounted for by a reduction factor that renormalizes the Gr\"uneisen parameter $\beta$ \cite{Woods2000,Suzuura2002,DeBeule2025}. This is a good approximation when out-of-plane displacements vary slowly with respect to the atomic lattice, which is our regime of interest.

\subsection{Intuitive picture and breakdown of scaling}

To illustrate the scaling law graphically, we show a scaled lattice in Fig.\ \ref{fig schematics}(e) \textit{without} scaling the displacements. Relative to the scaled lattice spacing, the unscaled displacement field is reduced, giving an intuitive picture of why $\mathbf{u}$ needs to be scaled to achieve the same PMF. A quantitative example of the scaling transformation is shown in Fig.\ \ref{fig schematics}(f), where we compare the PMF induced by triaxial strain on an unscaled and scaled graphene lattice based on Eq.\ \eqref{eq:scaling}, for small strains. Here, the strain is modeled with continuum elasticity. 

For slowly varying displacements, i.e., elastic strain fields, we find that the PMFs for the unscaled and scaled lattices are identical. However, when the displacements become too large, the PMF is not invariant under scaling, as shown in Fig.\ \ref{fig schematics}(g). Breakdown of the scaling law occurs when corrections to Eq.\ \eqref{eq:dt} for $\delta t_n$ that are quadratic in the strain tensor become significant. We estimate that scaling starts to break down for $su_{ij} \gtrsim 0.1$. (The strain tensor $u_{ij}$ is defined in Appendix~\ref{appendix pseudogauge fields}.) All calculations in this work are safely below this bound. Further confirmation of the applicability of the scaling law for in-plane displacements was obtained by studying pseudo-Landau levels (LLs) in a strained graphene flake, which is presented in the next section.

\section{Numerics}\label{sec numerics}

In this section, we numerically confirm the validity of the generalized scalable TBM with scaling law from Eq.\ \eqref{eq:scaling}. To this end, we performed extensive simulations of the site-resolved PMF, the local density of states (LDoS), and quantum transport by computing the conductance.

\subsection{Lattice PMF}\label{sec lattice pmf}

Although the main point of discussing the pseudogauge field in Section \ref{sec scalable TBM for strained graphene} is to understand how the displacement field should scale with the scaling factor $s$ in the scalable TBM, the explicit expression for $\mathbf{A}_s$ given by Eq.\ \eqref{eq As final} allows us to numerically compute the site-resolved PMF, $\mathbf{B}_s=\nabla\times\mathbf{A}_s$, without further approximation. This is how Figs.\ \ref{fig schematics}(f)--(g) were done, for which we explain more details here.

The hexagonal graphene flake of edge $D_h$ is first treated as a continuous medium and triaxially strained, as schematically shown in Fig.\ \ref{fig schematics}(c), where the outward displacement $\Delta$ applied on the three edges at $30^\circ,150^\circ,270^\circ$ relative to the $x$-axis enters the linear elasticity problem as boundary conditions. The shape of the deformed graphene flake is numerically obtained by \textsc{Matlab}'s Partial Differential Equation (PDE) Toolbox \cite{pdetoolbox}, using the Poisson ratio $\nu=0.165$ taken from the literature \cite{Blakslee1970,Pereira2009}. Note that all three schematics shown in Fig.\ \ref{fig schematics}(a)--(c) exemplifying in-plane (but exaggerated) strain are based on the above-described PDE simulation, with the grayscale background giving the von Mises stress \cite{pdetoolbox}.

%for scattered interpolants for obtaining
The finite-element mesh carrying the PDE solution for the deformed graphene flake is subsequently interpolated to obtain the position of the graphene lattice sites, whether scaled or unscaled. With the positions of the graphene lattice, the hopping distances to neighbors, and hence the hopping strengths based on Eq.\ \eqref{eq tij} with $t_0$ scaled by Eq.\ \eqref{eq:scaling}, can be obtained. Subtracted by the unstrained hopping strength, $t_0/s$, the change in the hopping amplitudes $\delta t_n$, can be used to obtain the pseudo vector potential $\mathbf{A}_s$ through Eq.\ \eqref{eq As final}. Finally, performing a numerical curl of $\mathbf{A}_s$ results in the site-resolved PMF $\mathbf{B}_s$ \cite{DeBeule2025}. The PMF profiles, $|\mathbf{B}_s|$, shown in Figs.\ \ref{fig schematics}(f)--(g) are based on the above-described procedure. 

Specifically, the left panel of Fig.\ \ref{fig schematics}(f) shows the lattice PMF profile based on an unscaled graphene hexagon of edge $D_h=20\unit{nm}$ undergoing a triaxial strain of $\Delta=0.1\unit{nm}\equiv \Delta_0$. Close to the center of the hexagon, the PMF profile is approximately uniform, consistent with Ref.\ \onlinecite{Guinea2009}. The almost identical PMF profile is obtained in the $s=2$ scaled graphene lattice, subject to $\Delta=2\Delta_0=0.2\unit{nm}$ based on the scaling summarized in Eq.\ \eqref{eq:scaling}. However, the scaling of the displacement may fail when the displacement field is too strong, as exemplified in Fig.\ \ref{fig schematics}(g), where an unscaled $\Delta_0=2.5\unit{nm}$ was considered.

\subsection{Local density of states}\label{sec ldos}

The tight-binding Hamiltonian for a finite-size graphene lattice based on Eq.\ \eqref{eq:Htb} is generally an $N\times N$ matrix, where $N$ is the total number of lattice sites. The Landau levels due to either external magnetic field $B_z$, pseudomagnetic field $B_s$, or their combination, can be visualized by inspecting the local density of states (LDoS). For a lattice site $n$ and energy $E$, the LDoS can be obtained from the imaginary part of the retarded Green's function $G^R(E) = \left[ (E+i\eta) \openone_N - H \right]^{-1}$, where $\eta$ introduces a phenomenological broadening and $\openone_N$ is the $N\times N$ identity matrix, as
\begin{equation}
    D_n(E) = -\frac{1}{\pi}\Im G^R_{n,n}(E)\ . \label{eq ldos}
\end{equation}
In all LDoS calculations presented in this work, we set $\eta = 5\unit{meV}$.

% slightly changed here based on Aitor's comments

%where $i$ labels the sites. This is the imaginary part of the $i$th diagonal matrix element of the retarded Green's function $G^R(E)$, also an $N\times N$ matrix which is the inverse of $(E+i\eta)\openone - H$. Here, $\openone$ is the $N\times N$ identity matrix and $i\eta$ is a small imaginary number that gives a phenomenological broadening of the energy levels, set to be $\eta = 5\unit{meV}$ in the following calculations.

\begin{figure}[t]
\includegraphics[width=\columnwidth]{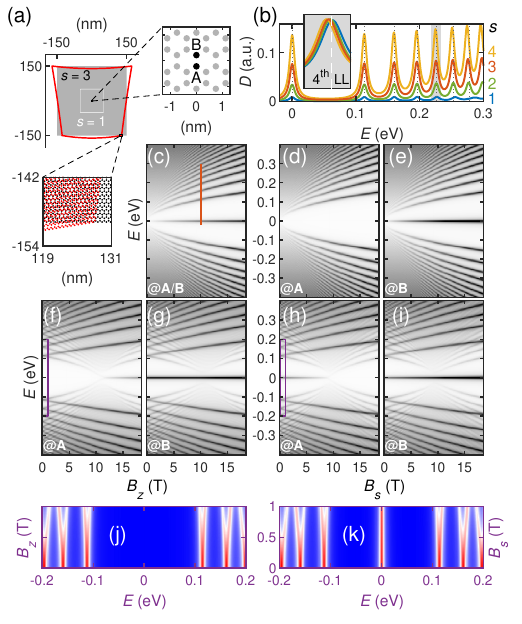}
\caption{Local density of states at the center of a scaled and strained zigzag graphene flake, $D$. Shown in (a) for $s=3$ without strain (gray square) and with strain (deformed red square). The corresponding lattice sites are shown in the bottom inset. (b) $D(E)$ for unstrained graphene scaled by $s=1,2,3,4$ for an external magnetic field $B_z=10\unit{T}$. The inset shows $D/s^2$ near the 4th LL. The red curve in (b) for $s=3$ gives the line cut marked in (c), showing identical $D(B_z,E)$ on both sublattices [defined in the right inset of (a)]. Panels (d) and (e) show $D(B_s,E)$ at sites A and B, respectively, for $B_z=0$ and the same range of the PMF $B_s$ as for $B_z$ in (c). Panels (f) and (g) show $D(B_z,E)$ at site A and B, respectively, for $B_s=10\unit{T}$. Panels (h) and (i) show $D(B_s,E)$ at site A and B, respectively, for $B_z=10\unit{T}$. Panels (j) and (k) are enlarged (and rotated) low-field maps marked by the purple boxes on (f) and (h), respectively. Here $D \in  [0,0.15]$ for all maps such that white [blue] and black [red] correspond to zero and the maximum, respectively, in panels (c)--(i) [(j)--(k)].}
\label{fig LDoS}
\end{figure}

\subsubsection{Landau and pseudo-Landau levels}

We now further benchmark the scalable TBM and show the scaling invariance of the LLs and pseudo-LLs. To this end, we consider a scalable lattice of dimensions $400 \times 231$ hexagons, corresponding roughly to an area of $50s \times 50s \,\unit{nm^2}$ as schematically shown in Fig.\ \ref{fig LDoS}(a), subject to an applied magnetic field $B_z$ perpendicular to the graphene plane and an in-plane strain given by the following displacement field:
\begin{equation}
    \mathbf{u}(x,y) = c\begin{pmatrix}
        2xy \\
        x^2-y^2
    \end{pmatrix}\ ,
    \label{eq u for constant Bs}
\end{equation}
where the constant $c$ is a control parameter in units of inverse length. The above displacement field \eqref{eq u for constant Bs} is expected to generate a constant pseudomagnetic field
\begin{equation} \label{eq:Bs}
    B_s = \frac{4\hbar\beta}{ea_0}c
\end{equation}
% Christophe: Ming-Hao I computed this again independently and I obtain this equation with a_0 the nn distance. You see that our scaling law effectively sends c -> sc and a_0 -> sa_0. I added a sentence to explain this
in the nearest-neighbor approximation \cite{Guinea2009}.
%in the nearest-neighbor approximation, similar to Ref.\ \onlinecite{Guinea2009} except that $a=sa_0$ here, instead of $a_0$.
As promised, we see that the expression for the PMF in Eq.\ \eqref{eq:Bs} is left invariant by the scaling law, given in Eq.\ \eqref{eq:scaling}, which effectively sends $c \rightarrow s c$ and $a_0 \rightarrow sa_0$. We now calculate the LDoS as a function of energy $E$ at the two central sites of the flake (belonging to different sublattices) marked on Fig.\ \ref{fig LDoS}(a), and which is denoted by $D$ for simplicity.

We start with the LLs due to $B_z$ in the absence of strain, shown in Fig.\ \ref{fig LDoS}(b) for $B_z=10\unit{T}$ and different scaling $s=1,2,3,4$. We see that the energy of the LLs is unchanged, but the peak height scales with the area as $s^2$ because it is the LDoS per unit area $\propto D(E)/s^2$ that is scaling invariant, rather than the LDoS. This also means that, whenever applying our scaling method to investigate lattice-site-resolved quantities, care must be taken due to the changing unit-cell area. Examples of $D(E)/s^2$ are shown in the inset of Fig.\ \ref{fig LDoS}(b) near the 4th LL, indicating that the peaks are slightly shifted for different $s$. We attribute this to finite-size effects that are more pronounced for smaller $s$, as the LLs become more extended for increasing $|n|$. As well as the reduced energy window corresponding to the Dirac regime with increasing $s$. Closer to charge neutrality, this shift becomes negligible on the meV scale. Further LDoS calculations to confirm the $s$ invariance are shown in Appendix \ref{appendix LDoS scaling invariance}.

We now focus on $s=3$ and show $D$ as a function of not only $E$ but also either $B_z$ with $B_s$ fixed or $B_s$ with $B_z$ fixed. The red curve in Fig.\ \ref{fig LDoS}(b) corresponds to the vertical line cut in Fig.\ \ref{fig LDoS}(c) for $D(B_z,E)$ with $B_s=0$, where the dark lines give LDoS peaks, perfectly matching the bulk LL energies:
\begin{equation}
    E_m(B) = \sgn(m)\sqrt{2e|B|\hbar v_F^2|m|}\ ,\quad m=0,\pm 1,\pm 2,\ldots\ .
\label{eq LL}
\end{equation}
Note that Eq.\ \eqref{eq LL} applies to the general case of coexisting real and pseudomagnetic fields with $B = B_z + \tau B_s$. In Figs.\ \ref{fig LDoS}(d) and (e), we show $D(B_s,E)$ with $B_z=0$ on the sites of sublattices A and B at the center of the flake, respectively. This reveals pseudo-LLs $|m|\geq 1$ on A and all $m$ on B. This is because the zeroth pseudo-LL only has support on one sublattice for both valleys, as explained in Appendix \ref{appendix pseudo LL}. We consider $0 \leq c \leq 3s\times 10^{-4}\unit{nm^{-1}}$ that leads to a $B_s$ in approximately the same range as in Fig.\ \ref{fig LDoS}(c) with $0\leq B_z\leq 18\unit{T}$. 

\subsubsection{Coexisting Landau and pseudo-Landau levels}

Next, we consider co-existing $B_z$ and $B_s$. In Figs.\ \ref{fig LDoS}(f)--(g), we show $D(B_z,E)$ on the central A and B site, respectively, for a strained graphene lattice with $c = 4.88\times 10^{-4}\unit{nm^{-1}}$ corresponding to $B_s = 10\unit{T}$. Due to the opposite $B_s$ for the two valleys, the LDoS reveals the superposed spectra $E_m(B)$ of Eq.\ \eqref{eq LL} with $B=B_z \pm B_s$. Without disturbing Figs.\ \ref{fig LDoS}(f) and \ref{fig LDoS}(g), the perfect match of the LDoS plots with $E_m(B_z\pm 10 \unit{T})$ is shown in Fig.\ \ref{fig LDoS cf formula} of Appendix \ref{appendix LL: LDoS vs formula}.

% As seen in Figs.\ \ref{fig LDoS}(f) and \ref{fig LDoS}(g), the LLs of one valley increase monotonically in energy, while those of the other valley decrease until $B_z = B_s$. We also note that the 0th LL only has support on the A sublattice for $B > 0$. At this point, one valley shifts its support to the other sublattice. 
As seen in Figs.\ \ref{fig LDoS}(f) and \ref{fig LDoS}(g), the LLs of one valley increase monotonically in energy, while those of the other valley decrease until $B_z = B_s$ when the field vanishes in one valley ($\tau = -1$ for the figures). We also note that the 0th LL has finite support on the A sublattice only for $\tau B < 0$ (see Appendix \ref{appendix pseudo LL}), which is the case for valley $\tau = -1$ when $B_z > B_s$ in both figures. Hence, at $B_z = B_s$, the effective total field B changes sign for valley $\tau = -1$ and shifts its support to the opposite sublattice. This is not clearly visible in Fig.\ \ref{fig LDoS}(g) because the nascent LLs for $B_z \gtrsim B_s$ are subject to finite-size effects. Similarly, by fixing the external magnetic field at $B_z=10\unit{T}$ and varying the strain, we obtain Fig.\ \ref{fig LDoS}(h) for site A and Fig.\ \ref{fig LDoS}(i) for site B. We further observe splitting of the (pseudo) LLs in the presence of a strong ($B_s$) $B_z$ when a weak ($B_z$) $B_s$ is applied. This is seen in Figs.\ \ref{fig LDoS}(j) and \ref{fig LDoS}(k), which are the enlarged maps marked by the purple box on Figs.\ \ref{fig LDoS}(f) and \ref{fig LDoS}(h), respectively. The latter has been discussed in Ref.\ \onlinecite{Li2020a}. 

Systematic calculations of $D(B_z,E)$ for fixed $B_s = 1,2,\ldots,15\unit{T}$ are shown in Appendix \ref{appendix coexisting LLs and pseudo-LLs at various Bs}.

\begin{figure}[t]
\includegraphics[width=\columnwidth]{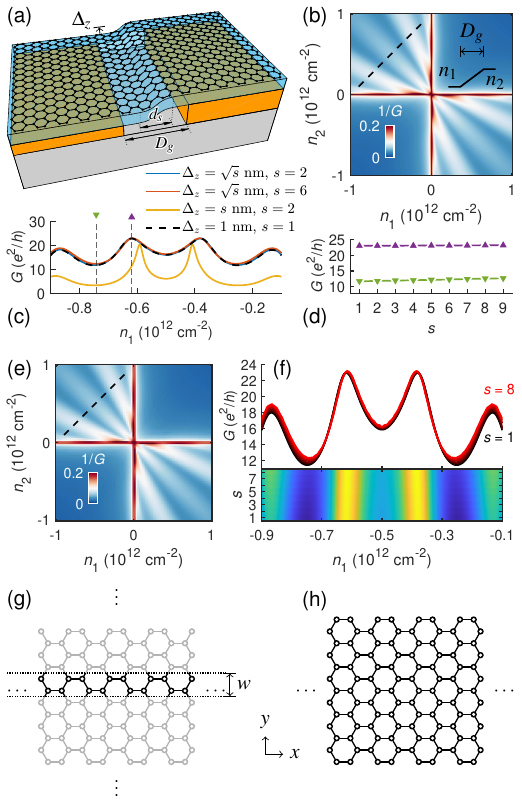}
\caption{Transport simulations for a graphene nanoslide \cite{Zhang2022,DeBeule2026} illustrated in (a), using (b)--(d) periodic and (e)--(f) open boundary conditions. The width of the ribbon is $W=500\unit{nm}$. (b) Inverse conductance $1/G$ versus carrier densities $n_1$ and $n_2$ in the left and right leads, respectively, for an unscaled ($s=1$) lattice. Inset: Carrier density profile used in the transport simulations. The conductance along the dashed line is shown in (c) and compared to results with properly and improperly scaled displacements. This illustrates the square-root scaling of the out-of-plane displacement field (see the legend). (d) Scaling factor dependence of the conductance with correct scaling at the carrier densities marked by the vertical dashed lines in (c). (e) Inverse conductance map for open boundary conditions computed with $(s,\Delta_z)=(4,2\unit{nm})$. (f) shows $G$ along the dashed line in (e) with $(s,\Delta_z)=(1,\sqrt{1}\unit{nm}),(2,\sqrt{2}\unit{nm}),\cdots,(8,\sqrt{8}\unit{nm})$, presented as individual curves (upper part) and a color map (lower part). Schematics showing lattices under the (g) periodic and (h) open boundary conditions along the $y$ direction.}
\label{fig nanoslide}
\end{figure}

\subsection{Simulations revisiting an experiment on transport through a uniaxial strain barrier}\label{sec transport}

%So far, we have implemented the generalized scalable TBM to compute the lattice PMF and the LDoS of strained graphene, confirming the scaling invariance of the in-plane component of the displacement field. 

% Aitor suggested not to use "so far" for being a casual tone

Having confirmed the scaling invariance of the in-plane component of the displacement field by showing the lattice PMF and the LDoS, we continue to confirm the square root scaling behavior of the out-of-plane component summarized in Eq.\ \eqref{eq:scaling} by performing quantum transport simulations. To this end, we revisit a recent experiment on a two-terminal graphene p-n junction tunable by two local gates with a thickness difference of about 8~nm \cite{Zhang2022}, leading to a local strain barrier, as schematically shown in Fig.\ \ref{fig nanoslide}(a).

The intentional thickness difference causes a vertical misalignment of the graphene flake between the left and right regions, forming a ``graphene nanoslide'' modeled by the following height profile:
\begin{equation}
    h(x) = \frac{\Delta_z}{2} \, \tanh \left( \frac{4x}{d_s} \right), 
    \label{eq h(x)}
\end{equation}
where $\Delta_z$ is the height difference and $d_s$ sets the length scale over which the height changes; see Fig.\ \ref{fig nanoslide}(a). To avoid complications arising from the gate geometry, we model the net carrier density profile induced by the two local gates with a simple linear function shown in the inset of Fig.\ \ref{fig nanoslide}(b).

Here, we do not expound the rich underlying physics of the graphene nanoslide device (such as why the lattice direction is crucial and why there are conductance oscillations in the bipolar regime), which can be found in Ref.\ \onlinecite{DeBeule2026}. Instead, we focus only on confirming the scaling behavior of the vertical displacements, as summarized in Eq.\ \eqref{eq:scaling}, by performing quantum transport simulations with various scaling factors, considering both infinitely wide and finite-width graphene ribbons.

\subsubsection{Periodic boundary conditions}

% \times W/(\sqrt{3}\pi sa)
% we define a_0 as the carbon-carbon distance in the intro
The height profile, Eq.\ \eqref{eq h(x)}, as well as the gate-controlled carrier density profile, $n(x)$, do not depend on $y$. Hence, assuming translation invariance along $y$, the conductance based on the Landauer formula can be well approximated by
\begin{equation}
    G = G_0\sum_{k_y} T(k_y) \rightarrow \frac{G_0}{2\pi/W} \int T(k_y)dk_y = \frac{G_0 W}{2\pi w} g,
    \label{eq G Landauer with PBC}
\end{equation}
where $G_0=2e^2/h$ is the conductance quantum taking into account spin, $w$ is the unit cell width along $y$ (for armchair and zigzag along $x$, we have $w=\sqrt{3}s a_0$ and $w=3sa_0$, respectively), and $g=\int T(\varphi)d\varphi$ can be computed using the real-space Green's function method \cite{Datta1995,Chakraborti2024} with the periodic hopping amplitudes $t \rightarrow t e^{\pm i\varphi}$ at the boundary \cite{Wimmer2008,Chakraborti2024}. Since $T(\varphi)$ is periodic in the Bloch phase $\varphi = k_y w$ with period $2\pi$, the computation of $g$ in Eq.\ \eqref{eq G Landauer with PBC} can be carried out by numerically integrating $T(\varphi)$ over any interval of length $2\pi$.
%, such as $\varphi \in [-\pi,\pi]$ or $\varphi \in [0,2\pi]$.

%Hence, we can use the translational invariance to define a transverse momentum $k_y$ with appropriate Bloch phases for the hopping amplitudes at the boundary \cite{Wimmer2008,Chakraborti2024}. From the integrated transmission function, $g = \sum_{k_y} T(k_y) \rightarrow W/(2\pi) \int T(k_y) dk_y$, where $T$ is computed with the real-space Green's function method \cite{Datta1995,Chakraborti2024}, we obtain the conductance for width $W \gg D_g$ via $G = (2e^2/h) \times g$. Here the factor $2$ accounts for spin and the width $W$ sets the momentum spacing: $k_y = 2\pi m/W$ where $m = 0,\ldots,N-1$ with $N = W/(3sa_0)$.

While the parameters that best fit the experiment of Ref.\ \onlinecite{Zhang2022} are $(D_g,d_s,\Delta_z)\approx (50,24,8)\unit{nm}$, here we adopt $(D_g,d_s,\Delta_z)=(100,10,1)\unit{nm}$, to account for the gate geometry and in-plane relaxation, phenomenologically. Taking $W=500\unit{nm}$ and adopting an unscaled graphene lattice, the inverse of the conductance as a function of $n_1$ and $n_2$ is shown in Fig.\ \ref{fig nanoslide}(b), qualitatively agreeing with the resistance map reported in Ref.\ \onlinecite{Zhang2022}. The conductance trace marked by the black dashed line shown in Fig.\ \ref{fig nanoslide}(c) is further compared with other calculations using scaled graphene lattices. As clearly seen, results based on $(s,\Delta_z)=(2,\sqrt{2}\unit{nm})$ and $(s,\Delta_z)=(6,\sqrt{6}\unit{nm})$ nearly perfectly overlap with the unscaled result of $(s,\Delta_z)=(1,1\unit{nm})$. However, when the vertical component of the displacement is not properly scaled, the result is very different, as exemplified by $(s,\Delta_z)=(2,2\unit{nm})$ curve shown in Fig.\ \ref{fig nanoslide}(c). The values of the conductance dip and peak, considering properly scaled $\Delta_z$ marked by the triangles in Fig.\ \ref{fig nanoslide}(c), are plotted in Fig.\ \ref{fig nanoslide}(d) against the scaling factor $s$, confirming the scaling invariance.

\subsubsection{Open boundary conditions}

Alternatively, modeling the $500$-nm-wide graphene as a finite-width armchair ribbon with open boundary conditions (i.e., the lattice is terminated at $y=\pm W/2$), we perform quantum transport simulations using the open-source \textsc{Python} package, \textsc{Kwant} \cite{Groth2014}. The inverse conductance map based on a graphene ribbon scaled by $s=4$ and strained by $\Delta_z=2\unit{nm}$ is shown in Fig.\ \ref{fig nanoslide}(e), which is hardly distinguishable from the map of Fig.\ \ref{fig nanoslide}(b) based on the periodic boundary hopping method. Line cuts along the black dashed line marked on Fig.\ \ref{fig nanoslide}(e) are compared in Fig.\ \ref{fig nanoslide}(f) for $(s,\Delta_z)=(1,\sqrt{1}\unit{nm}),(2,\sqrt{2}\unit{nm}),\cdots,(8,\sqrt{8}\unit{nm})$ presented as curves (upper part of the panel) and a color map (lower part of the panel), confirming again the scaling law for the out-of-plane displacement field.

\subsubsection{Remarks on boundary conditions}

For clarity, the two types of lattices under the periodic and open boundary conditions are schematically illustrated in Fig.\ \ref{fig nanoslide}(g) and Fig.\ \ref{fig nanoslide}(h), respectively, where centered horizontal ellipses indicate that the lattice extends to $x\rightarrow\pm\infty$. The vertical ellipses in Fig.\ \ref{fig nanoslide}(g) imply that the lattice is assumed to extend to $y\rightarrow\pm \infty$. Specifically, transport simulations shown in Figs.\ \ref{fig nanoslide}(b)--(d) are based on the unit cell highlighted in Fig.\ \ref{fig nanoslide}(g), and those shown in Figs.\ \ref{fig nanoslide}(e)--(f) are based on the finite-width lattice shown in Fig.\ \ref{fig nanoslide}(h).

Note that the transport simulations presented here enable a rigorous comparison between scaled and unscaled lattices under both periodic and open boundary conditions. This is possible because of the geometry of the revisited experiment of the graphene nanoslide and the assumed width of $W=0.5\unit{\mu m}$. In most transport experiments, periodic boundary conditions are not applicable either because of the device geometry (e.g., Hall-bar measurements) or because of the presence of strong magnetic fields, where edges are essential. In such cases, finite-width calculations with open boundary conditions are unavoidable. Typical graphene samples used in experiments have dimensions on the order of $1\unit{\mu m}$ in both length and width, making transport simulations based on unscaled lattices computationally infeasible. 

% no abbrevations in the summary
\section{Summary}\label{sec summary}

We have shown that the scalable tight-binding model \cite{Liu2015} is readily applicable to strained graphene, provided that the displacement fields are properly scaled following Eq.\ \eqref{eq:scaling}. Our theory holds as long as the scaled displacement fields vary slowly with respect to the scaled lattice, such that the scaled strain tensor remains in the elastic regime. Moreover, in this case, the scaled model still faithfully reproduces the long-wavelength Dirac theory in the presence of strain-induced pseudogauge fields. To confirm the scaling invariance of the generalized scalable tight-binding model, we performed systematic numerical calculations of the lattice pseudomagnetic field, as well as the local density of states. The latter was used to study pseudo-Landau levels and their competition with real Landau levels due to an external magnetic field, as well as their mixture defined by an effective valley-dependent magnetic field. In addition, we also performed quantum transport simulations to verify the scaling of out-of-plane displacements, taking the recent experiment on transport across a uniaxial strain barrier as a realistic example \cite{Zhang2022}.

We believe our work paves the way towards large-scale quantum transport simulations for strained graphene in the elastic long-wavelength regime, allowing for revisiting more experiments on strained graphene \cite{Levy2010,Lee2013,Gutierrez2016,Jia2019,Nigge2019,Mao2020,Li2020a,Wang2020,Ho2021,Zhou2023,Kapfer2023,Zhou2023,Kerjouan2024,McRae2024,SrutRakic2024,Sahani2026, Jiang2017} as well as previous interesting theoretical predictions \cite{Guinea2009,Guinea2010,Low2011,Moldovan2013,Zhu2015,Settnes2016,Settnes2016a,Settnes2016,Milovanovic2016,Cavalcante2016,CarrilloBastos2016,Salerno2017,Stegmann2018,LantagneHurtubise2020,Milovanovic2020,Zuo2022,Li2023}. Our work will facilitate the modeling of mesoscopic strained devices, thereby advancing the field of graphene straintronics.

\begin{acknowledgments}

We thank L.\ Wang, P.\ Makk, and C.\ Sch\"onenberger for illuminating discussions that greatly inspired this work. M.-H.L., H.-Y.W., and A.G.-R.\ acknowledge the National Center for High-performance Computing (NCHC) for providing computational and storage resources, and the National Science and Technology Council (NSTC) of Taiwan (Grant No.\ 112-2112-M-006-019-MY3, No.\ 113-2918-I-006-004, No.\ 114-2811-M-006-038, and No.\ 114-2112-M-006-029-MY3) for financial support. C.D.B.\ was supported by the U.S.\ Department of Energy under Grant No.\ DE-FG02-84ER45118. A.M.-K.\ acknowledges partial support by the program ``Excellence initiative -- research university'' for the AGH University of Krakow, and by Polish high-performance computing infrastructure PLGrid (HPC Center: ACK Cyfronet AGH) for providing computer facilities and support within computational grant No.\ PLG/2024/017407. D.K.~acknowledges NSTC of Taiwan under Grant No.~114-2112-M-006-034-MY3 for financial support. K.R.\ acknowledges funding through the Deutsche Forschungsgemeinschaft (DFG, German Research Foundation) within Project-ID 314695032--SFB 1277.

%We thank L.\ Wang, P.\ Makk, and C.\ Sch\"onenberger for illuminating discussions that greatly inspired this work. A.M.-K.\ acknowledges partial support by the program ``Excellence initiative -- research university'' for the AGH University of Krakow, and by the Polish high-performance computing infrastructure PLGrid (HPC Center: ACK Cyfronet AGH) for providing computer facilities and support within computational grant no.\ PLG/2024/017407. C.D.B.\ was supported by the U.S.\ Department of Energy under Grant No.\ DE-FG02-84ER45118. J.-T.S, A.G.-R., and M.-H.L.\ acknowledge the National Center for High-performance Computing (NCHC) for providing computational and storage resources, and the National Science and Technology Council (NSTC) of Taiwan (Grant No.\ 112-2112-M-006-019-MY3, 113-2918-I-006-004, 113-2811-M-006-035) for financial support. D.K.~acknowledges partial support from the project IM-2021-26 (SUPERSPIN) funded by the Slovak Academy of Sciences via the program IMPULZ. K.R.\ acknowledges funding through the Deutsche Forschungsgemeinschaft (DFG, German Research Foundation) within Project-ID 314695032 -- SFB 1277. It furthermore supported, together with the Alumni Program of the Alexander von Humboldt Foundation, M.-H.L.'s sabbatical stay at the University of Regensburg, where part of this work was done.
\end{acknowledgments}

\appendix

\begin{figure*}[t]
\subfigure[\label{fig LDoS vs Bz for various s}]{
\includegraphics[width=\textwidth]{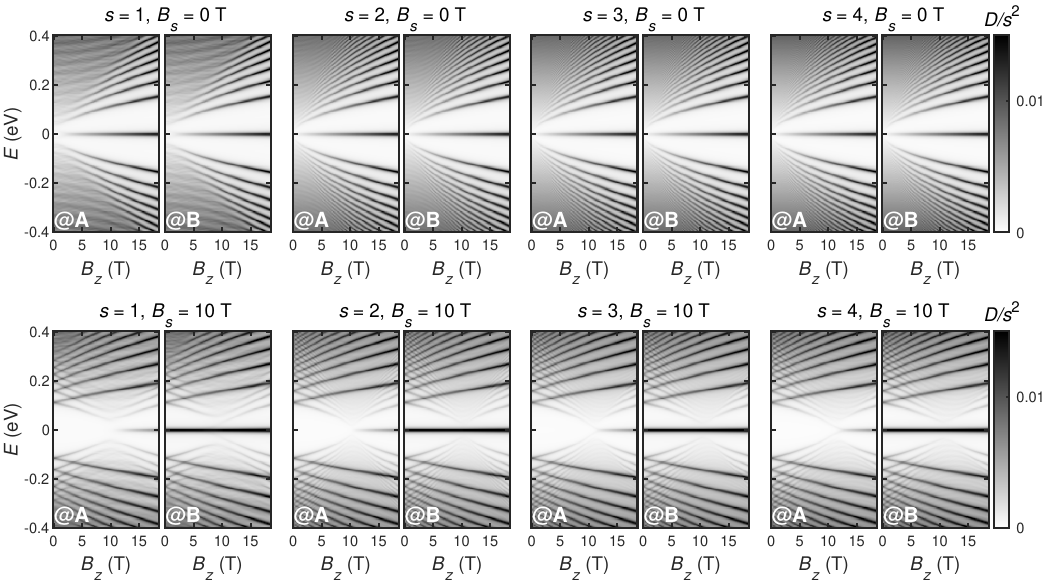}
}

\subfigure[\label{fig LDoS vs Bs for various s}]{
\includegraphics[width=\textwidth]{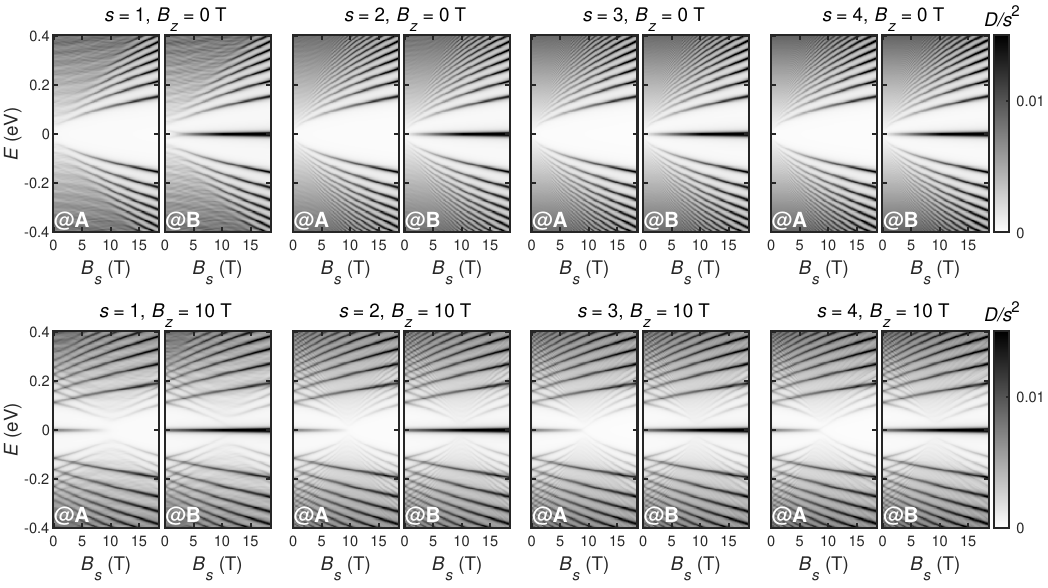}
}
\caption{(a) Local density of states normalized by the squared scaling factor, $D/s^2$, as a function of the external magnetic field $B_z$ at zero pseudomagnetic field $B_s=0$ (upper row) and finite pseudomagnetic field $B_s=10\unit{T}$ (lower row). (b) Same as (a) but with $B_z$ and $B_s$ interchanged.}
\end{figure*}

\section{Additional LDoS calculations}\label{appendix additional data}

In Fig.\ \ref{fig LDoS}, we have shown how the LDoS of a graphene lattice varies with the energy $E$, external magnetic field $B_z$, in-plane strain following Eq.\ \eqref{eq u for constant Bs} that leads to a constant pseudomagnetic field $B_s$, the scaling factor $s$, and the sublattice. At a strong $B_z$ with $B_s=0$, LLs are seen. At strong $B_s$ with $B_z=0$, pseudo-LLs are seen. With finite $B_z$ and $B_s$, coexisting LLs and pseudo-LLs are seen. Here, we elaborate with more LDoS plots to carefully confirm the scaling invariance of the scalable TBM described by Eq.\ \eqref{eq:scaling} as well as how well the revealed LLs and pseudo-LLs agree with Eq.\ \eqref{eq LL}.

\subsection{Scaling invariance}\label{appendix LDoS scaling invariance}

In Fig.\ \ref{fig LDoS}(c), the LLs are revealed by the LDoS as functions of $B_z$ and $E$ using a graphene lattice scaled by $s=3$ free of strain. In the upper row of Fig.\ \ref{fig LDoS vs Bz for various s}, the same LLs for various scaling factors ($s=1,2,3,4$) can be seen for both A and B sites. In the lower row of Fig.\ \ref{fig LDoS vs Bz for various s}, similar plots but with a fixed strength of in-plane strain corresponding to $B_s=10\unit{T}$ are shown, confirming the scaling invariance of the local density of states in the scalable TBM.

Similarly, the pseudo-LLs are revealed by the LDoS as functions of $B_s$ and $E$ in Fig.\ \ref{fig LDoS}(d) and (e), using a graphene lattice scaled by $s=3$, free of external magnetic field. In the upper row of Fig.\ \ref{fig LDoS vs Bs for various s}, the same pseudo-LLs for various scaling factors ($s=1,2,3,4$) can be seen for both A and B sites. In the lower row of Fig.\ \ref{fig LDoS vs Bs for various s}, similar plots but with a fixed strength of external magnetic field $B_z=10\unit{T}$ are shown.
\begin{figure}
    \includegraphics[width=\columnwidth]{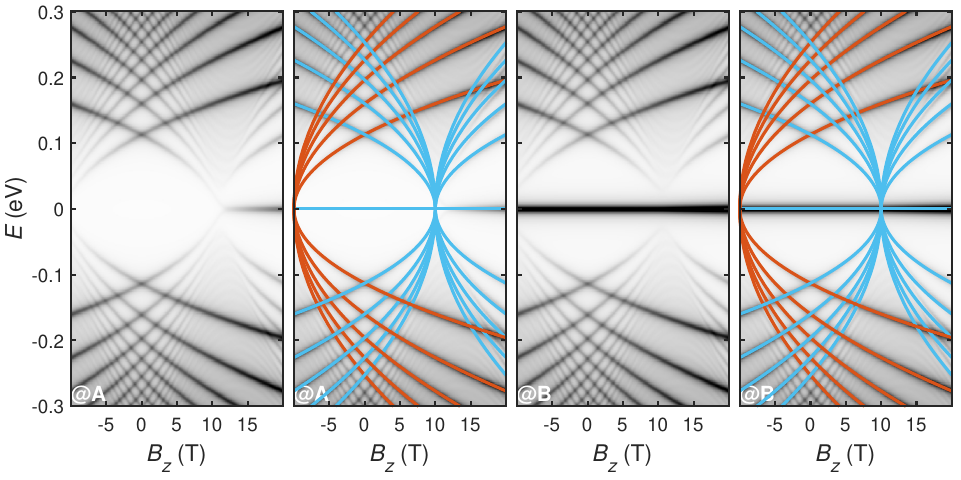}
    \caption{Local density of states versus the external magnetic field $B_z$ with a constant PMF $B_s=10\unit{T}$ on the center A site (left two panels) and B site (right two panels) compared to the Landau level formula of Eq.\ \eqref{eq LL}. For clarity, the LDoS maps for each sublattice are shown both with and without superimposing the bulk Landau levels $E_m(B_z+10\unit{T})$ for valley $\mathbf K$ (red) and $E_m(B_z-10\unit{T})$ (cyan) for valley $\mathbf K'$ using Eq.\ \eqref{eq LL}.}
    \label{fig LDoS cf formula}
\end{figure}

\subsection{LLs and pseudo-LLs: Numerics vs analytics}\label{appendix LL: LDoS vs formula}

Figure \ref{fig LDoS cf formula} shows the same data from Figs.\ \ref{fig LDoS}(f) and \ref{fig LDoS}(g), which are LDoS as functions of $B_z$ at a fixed in-plane strain that corresponds to a uniform PMF of strength $B_s=10\unit{T}$, but with a slightly different plot range to better highlight the coexisting LLs and pseudo-LLs. The left two panels of Fig.\ \ref{fig LDoS cf formula} are for sublattice A corresponding to Fig.\ \ref{fig LDoS}(f), one without the curves given by $E_m(B_z\pm 10\unit{T})$ based on Eq.\ \eqref{eq LL} and one with. Similarly, the right two panels of Fig.\ \ref{fig LDoS cf formula} are for sublattice B, corresponding to Fig.\ \ref{fig LDoS}(g). Perfect agreement between the numerical LDoS evaluated at the central sites and the analytical formula can be seen for the lowest six Landau levels.

Note that the zeroth Landau level is localized entirely on sublattice B in both valleys as long as $\tau B = B_s + \tau B_z$ is positive, i.e., for $B_z \in [-|B_s|,|B_s|]$, consistent with the analytical result of Eq.\ \eqref{eq:LLwf}. Outside of this regime, e.g., for $B_z > |B_s|$, the zeroth Landau level in valley $\mathbf K$ ($\tau = +1$) is localized on sublattice B, while in valley $\mathbf K'$ ($\tau = -1$) it is localized on sublattice A. Here, the difference in the LDoS magnitude between the A and B sublattices is due to the fact that the total magnetic fields in different valleys have different magnitudes. One of them is close to zero, resulting in a density that is spread out much more compared to the other one, which already has a large total magnetic field and whose wave function is thus much more localized. In symmetric gauge, only the $L_z = 0$ mode contributes at the center of the strained graphene flake with $|\psi_{0,0}|^2 \propto e^{-r^2/(2\ell_B^2)}$.
\begin{figure*}
    \subfigure[\label{fig coexisting LL and pseudoLL at various Bs: A}]{
    \includegraphics[page=1,width=\textwidth]{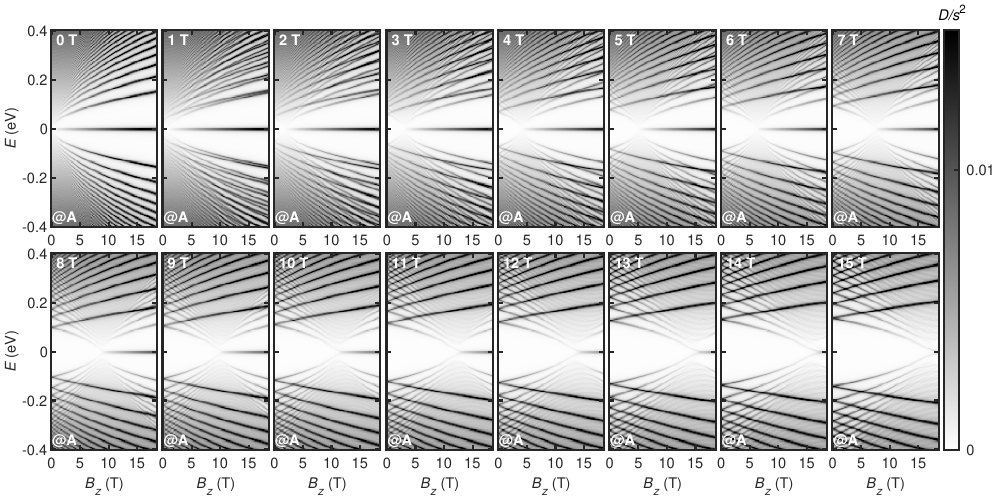}}
    \subfigure[\label{fig coexisting LL and pseudoLL at various Bs: B}]{
    \includegraphics[page=2,width=\textwidth]{fig_LDoS_vs_Bz_at_variousBs.pdf}}
    \caption{(a) Local density of states normalized by the squared scaling factor, $D/s^2$, with $s=3$ as a function of the external magnetic field $B_z$ at various pseudomagnetic field $B_s=0,1,\cdots,15\unit{T}$ (upper left corner of each panel) at the center A site. Results for the center B site are shown in (b). All panels share the same color range indicated by the color bars on the right-hand side.}
    \label{figS DoS for various Bs}
\end{figure*}

\subsection{Coexisting LLs and pseudo-LLs at various PMFs}\label{appendix coexisting LLs and pseudo-LLs at various Bs}

We have shown the coexisting LLs and pseudo-LLs, either for the case of varying $B_z$ at a fixed $B_s=10\unit{T}$, or for the case of varying $B_s$ at a fixed $B_z=10\unit{T}$. The former are shown in Figs.\ \ref{fig LDoS}(f)--(g), panels of the lower row of Fig.\ \ref{fig LDoS vs Bz for various s}, and Fig.\ \ref{fig LDoS cf formula}; the latter are shown in Figs.\ \ref{fig LDoS}(h)--(i) and panels of the lower row of Fig.\ \ref{fig LDoS vs Bs for various s}. Here, we elaborate the former case of $D(B_z,E)$ at various strengths of in-plane strain corresponding to $B_s=0,1,2,\cdots,15\unit{T}$, as shown in Fig.\ \ref{fig coexisting LL and pseudoLL at various Bs: A} for sublattice A and Fig.\ \ref{fig coexisting LL and pseudoLL at various Bs: B} for sublattice B. The two sets of Landau levels, $E_m(B_z+B_s)$ and $E_m(B_z-B_s)$, can be clearly seen.

\section{Pseudogauge fields in graphene}\label{appendix pseudogauge fields}

A brief overview of the pseudogauge field in graphene is given here. Consider the graphene lattice shown in Fig.\ \ref{fig graphene nn vectors}, with the $x$ axis aligned along the zigzag direction. The pseudogauge field $\mathbf A_\mathrm{s} = (A_{s,x},A_{s,y})$ for valley $\mathbf K$ is defined as \cite{Katsnelson2007,Vozmediano2010,Amorim2016}
\begin{equation} \label{eq:As}
    A_{s,x} - i A_{s,y} = -\frac{1}{ev_F} \sum_{\mathbf a} \delta t(\mathbf a + \mathbf d_1^0) e^{i \mathbf K \cdot (\mathbf a + \mathbf d_1^0)},
\end{equation}
where the sum runs over lattice vectors $\mathbf a$ of the unstrained graphene and $\mathbf d_1^0 = a_0 \hat{\mathbf{e}}_y$. The change of the hopping amplitude due to lattice deformations is given in lowest order by
\begin{align}
    \delta t(\mathbf d) & = t(\mathbf d') - t(\mathbf d) \notag\\
    & = \frac{\partial t}{\partial d_j} (d_j' - d_j) + \frac{1}{2} \frac{\partial^2 t}{\partial d_z^2} (d_z' - d_z)^2,
\end{align}
where $j$ only runs over in-plane coordinates $x$ and $y$, and the derivatives are evaluated for pristine graphene. The change in a bond vector $\mathbf d$ due to lattice deformations between sublattices $\sigma$ and $\sigma'$, is given by
\begin{equation}
\begin{aligned}
    \mathbf d' - \mathbf d & = \mathbf{u}_\sigma(\mathbf r + \mathbf d) -\mathbf{u}_{\sigma'}(\mathbf r) + [ h_\sigma(\mathbf r + \mathbf d) - h_{\sigma'}(\mathbf r) ] \hat{\mathbf{e}}_z \\
    & \approx ( \mathbf d \cdot \nabla ) ( \mathbf u + h \hat{\mathbf{e}}_z) \\
    & + ( \delta_{\sigma A} \delta_{\sigma'B} - \delta_{\sigma B} \delta_{\sigma'A} ) ( \mathbf v + w \hat{\mathbf{e}}_z ),
\end{aligned}
\end{equation}
where $\mathbf u$ and $h$ are the in-plane and out-of-plane acoustic displacement fields, and $\mathbf v$ and $w$ are the in-plane and out-of-plane optical displacement fields \cite{DeBeule2025}. In the following, we will not include the optical parts and instead account for their contributions by a reduction factor \cite{Woods2000,Suzuura2002}. This is expected to hold for slowly-varying out-of-plane displacements and effectively renormalizes the magnitude of the pseudogauge field by a constant factor. If we further assume that the hopping amplitude only depends on the bond length, we obtain
\begin{align}
    \delta t & = \frac{1}{d} \frac{\partial t}{\partial d} \left[ d_j (d_j'-d_j) + \frac{1}{2} (d_z' - d_z)^2 \right] \\
    & = \frac{1}{d} \frac{\partial t}{\partial d} \, d_i d_j \left[ \partial_i u_j + \frac{1}{2} (\partial_i h) (\partial_j h) \right]
     = \frac{1}{d} \frac{\partial t}{\partial d} \, d_i d_j u_{ij}, \label{eq:delta_t}
    %\textbf{\frac{t'(d)}{d} }
\end{align}
where the symmetric piece gives the strain tensor $u_{ij} = \left[ \partial_i u_j + \partial_j u_i + (\partial_i h) (\partial_j h) \right] / 2$ and we used that
\begin{align}
    \frac{\partial t}{\partial d_i} & = \frac{d_i}{d} \frac{\partial t}{\partial d}, \quad
    \frac{\partial t}{\partial d_z} = \frac{d_z}{d} \frac{\partial t}{\partial d}, \\
    \frac{\partial^2 t}{\partial d_z^2} & = \frac{1}{d} \frac{\partial t}{\partial d} - \frac{d_z^2}{d^3} \frac{\partial t}{\partial d} + \frac{d_z^2}{d^2} \frac{\partial t^2}{\partial d^2},
\end{align}
together with the fact that $d_z = 0$ for pristine graphene. From the $C_{3z}$ rotation symmetry of graphene, or explicitly from Eq.\ \eqref{eq:delta_t}, it follows for our choice of coordinates that
\cite{Kang2023}:
\begin{equation}
%\frac{(a_x+d_{1,x}^0)^2 t'(\mathbf a + \mathbf d_1^0)}{|\mathbf a + \mathbf d_1^0|}
    e v_F \mathbf A_\mathrm{s} = \left[ \sum_{\mathbf a} \left. \frac{x^2 t'(r)}{r} \right|_{\mathbf r = \mathbf a + \mathbf d_1^0} e^{i \mathbf K \cdot (\mathbf a + \mathbf d_1^0)} \right]
    \begin{pmatrix}
        u_{yy} - u_{xx} \\ u_{xy} + u_{yx}
    \end{pmatrix},
\end{equation}
where the prefactor is a constant with units of energy that depends on the details of the hopping function. For nearest-neighbor hopping, as is done in the main text, the prefactor becomes $3\beta t_0/4$. We note that the object on the right-hand side transforms as a $d$ orbital with angular momentum $-2$ \cite{Venderbos2016}. It transforms only as a vector if we restrict to the valley-preserving symmetries of graphene.

\section{(Pseudo) Landau levels in graphene}\label{appendix pseudo LL}

To explain the zeroth Landau level seen differently on the two sublattices, we review the theory of a Dirac particle on the $xy$ plane in a magnetic field $\mathbf B = B \hat{\mathbf{e}}_z$. The Hamiltonian is given by
\begin{equation}
    H_0 = \hbar v_F \begin{pmatrix} 0 & \pi_- \\ \pi_+ & 0 \end{pmatrix},
\end{equation}
where $\pi_\pm = k_\pm + (e/\hbar) A_\pm$ with $k_\pm = \tau k_x \pm i k_y$ and $A_\pm = \tau A_x \pm i A_y$ ($\tau = \pm1$ is the valley index). Here, we take $e>0$ and $\mathbf A = (A_x, A_y)$ is the total vector potential, including both real and pseudogauge fields, with $B = (\nabla \times \mathbf A)_z$ the total magnetic field. Now consider the commutator
\begin{equation}
    \begin{aligned}
        \left[\pi_+, \pi_-\right] & = [k_+, k_-] + \frac{e}{\hbar} \left( [k_+,A_-] + [A_+,k_-] \right) \\
        & + \left( \frac{e}{\hbar} \right)^2 [A_+,A_-].
    \end{aligned}
\end{equation}
The first and last terms vanish, and the middle term obeys
\begin{equation}
    [k_+,A_-]f = (-i\partial_+ A_-) f, \quad [A_+,k_-]f = (i\partial_-A_+) f,
\end{equation}
where $f = f(x,y)$ is a test function and $\partial_\pm = \tau \partial_x \pm i \partial_y$. Hence, we find
\begin{align}
    [\pi_+, \pi_-] & = i \frac{e}{\hbar} \left( \partial_- A_+ - \partial_+ A_- \right) \\
    & = -\frac{2e \tau}{\hbar} \left( \partial_x A_y - \partial_y A_x \right),
\end{align}
or $[\pi_-, \pi_+] = (2e\tau/\hbar) \left( \nabla \times \mathbf A \right)_z = 2eB\tau/\hbar$. We now define the magnetic length
\begin{equation}
    l = \sqrt{\frac{\hbar}{e|B|}},
\end{equation}
so that $[\pi_-, \pi_+] = \sgn(\tau B) 2/l^2$. This prompts us to introduce
\begin{equation}
a = \frac{l}{\sqrt{2}} \, \pi_{-t}, \qquad a^\dag = \frac{l}{\sqrt{2}} \, \pi_{+t},
\end{equation}
where $t = \sgn(\tau B)$ with $\tau B = B_s + \tau B_z$. These operators can be interpreted as ladder operators since $[a, a^\dagger] = 1$. The Hamiltonian can thus be written as
\begin{align} \label{eq:h0a}
    H_0 & = \hbar \omega_c \begin{pmatrix} 0 & a \\ a^\dagger & 0 \end{pmatrix}, \qquad \tau B > 0, \\
    H_0 & = \hbar \omega_c \begin{pmatrix} 0 & a^\dagger \\ a & 0 \end{pmatrix}, \qquad \tau B < 0,
\end{align}
where $\omega_c = \sqrt{2}v_F/l$. The eigenstates are given by
\begin{align}
    & \psi_{\lambda,m\neq0} = \frac{1}{\sqrt{2}} \begin{pmatrix} \left| m - 1 \right> \\ \lambda \left| m \right> \end{pmatrix}, \quad \psi_{m=0} \begin{pmatrix} 0 \\ \left| 0 \right> \end{pmatrix}, \quad \tau B > 0, \label{eq:LLwf} \\
    & \psi_{\lambda,m\neq0} = \frac{1}{\sqrt{2}} \begin{pmatrix} \lambda \left| m \right> \\ \left| m - 1 \right> \end{pmatrix}, \quad \psi_{m=0} \begin{pmatrix} \left| 0 \right> \\ 0 \end{pmatrix}, \quad \tau B < 0,
\end{align}
where $\lambda=\pm1$ and $m=1,2,\ldots$ with eigenvalues $E_{\lambda,m} = \lambda \hbar \omega_c \sqrt{m}$ and $E_0 = 0$. Here, each Landau level comes in four flavors given by valley and spin. Importantly, there is another quantum number that does not appear in the Hamiltonian, which gives the macroscopic degeneracy of each Landau level, namely the number of flux quanta threading the system, and which is related to the guiding center of semiclassical electron orbits in the magnetic field \cite{Goerbig2024}.

Hence, we see that for a pure pseudomagnetic field $\tau B = B_s$, the wave function of the zeroth Landau level is polarized on a single sublattice for both valleys.

\bibliography{refs}

\end{document}